\documentclass[a4paper,11pt]{article}
\usepackage{pos, amsmath, amsfonts, amsthm}
\usepackage{graphicx,scalerel}
\newcommand\sbullet[1][.5]{\mathbin{\ThisStyle{\vcenter{\hbox{%
					\scalebox{#1}{$\SavedStyle\bullet$}}}}}%
}

\title{Update on two-level sampling for glueball observables in quenched QCD}

\author*[a]{Lorenzo Barca}
\author[b]{Jacob Finkenrath}
\author[c]{Francesco Knechtli}
\author[d]{Michael Peardon}
\author[a]{Stefan Schaefer}
\author[c]{Juan Andr\'es Urrea-Ni\~{n}o}

\affiliation[a]{John von Neumann-Institut f\"ur Computing NIC, Deutsches Elektronen-Synchrotron DESY, Germany}
\affiliation[b]{Department of Theoretical Physics, European Organization for Nuclear Research, CERN, CH-1211 Geneva 23, Switzerland}
\affiliation[c]{Department of Physics, University of Wuppertal, Gaußstrasse 20, 42119 Germany}
\affiliation[d]{School of Mathematics, Trinity College Dublin, Ireland}

\emailAdd{lorenzo.barca@desy.de}

\abstract{We report our progress in combining a two-level sampling algorithm 
with distillation techniques for calculations of disconnected diagrams in quenched QCD. 
The simulations are performed on a single ensemble at $\beta=6.0$ and volume $V=16^3\times 64$, 
and at a pion mass of $m_\pi\approx 760~\mathrm{MeV}$.}

\FullConference{The 41st International Symposium on Lattice Field Theory (LATTICE2024)\\
 28 July - 3 August 2024\\
Liverpool, UK\\}

\begin{document}
\begin{flushright}
    \texttt{DESY-25-014\\CERN-TH-2025-025}
\end{flushright}
\maketitle

\section{Introduction}
One of the ubiquitous challenges in lattice QCD simulations is the signal-to-noise ratio (S/N) of correlation functions \cite{Parisi:1983ae, Lepage:1989hd}.
Particularly demanding, even in pure gauge theory, is the computation of the spectrum of glueballs, hypothetical particles composed predominantly of gluons. 
One major difficulty of these computations is the analysis of disconnected contributions:
while the signal decays exponentially with the distance between the operators, the statistical noise remains constant with standard sampling techniques.
Alternative sampling methods, like the multilevel algorithm \cite{Luscher:2001up}, can be used to reduce the error in a more efficient way than standard samplings.
In \cite{Barca:2023arw, Barca:2024fpc}, we demonstrate that a two-level sampling can reduce more efficiently the statistical error of pure gauge glueball two-point functions at large distances. At this lattice conference, similar algorithms have been adopted to study for the first time glueball scattering in Yang-Mills theory \cite{maxlatt2024}. In principle, the same ideas can be applied to study glueball structure quantities like glueball gravitational form factors \cite{Abbott:2024bre}, by adopting a three-level algorithm to estimate glueball three-point functions.

However, the inclusion of fermions makes the glueball calculations more demanding.
First, for sufficiently light quarks the glueballs are unstable bound states 
and a finite-volume formalism is required to reconstruct information of the resonance.
Due to this, the inclusion of multiparticle operators with large overlap to all glueball decay modes is necessary, and it can be quite expensive.
For instance, there are more than 10 decay modes observed in the experiments for the scalar glueball candidate $f_0(1500)$ \cite{ParticleDataGroup:2024cfk},
although the largest fraction is due to $2\pi$ and $4\pi$ modes.
In addition to this, a comprehensive lattice analysis must resolve accurately all the states in the spectrum 
which lie below or close to the glueball energy.

The second two-fold challenge to face when simulating QCD is the presence of quark propagators.
From one side, they are very demanding to compute on each gauge configuration; however,
advanced solvers based on Krylov space solvers can be used to compute the propagators at a moderate cost even at large volumes.
On the other side, the non-locality of the quark propagator hinders the application of multilevel algorithms,
because they depend on the values of the gauge fields over the full space-time. However, by factorizing the quark propagator in different regions, it is possible to make the fermionic observables amenable for multilevel integration.

In quenched QCD, an important step forward has been made in \cite{Ce:2016idq} to rewrite the quark propagator 
as a series of terms with a well defined hierarchical structure. This factorisation of the quark propagator enables
a two-level integration of the fermionic observables.
In full QCD, the application of multilevel integration requires more advanced techniques due to the presence of the fermion determinant as the fermionic weight.
A factorisation of the fermion determinant via a multiboson approximation \cite{Luscher:1993xx} 
makes the fermionic observable and fermionic weight amenable for a local integration,
as demonstrated in \cite{Ce:2016ajy, DallaBrida:2020cik}.

As a step towards the computation of the glueball spectrum in full QCD, we combine for the first time distillation techniques
with the multilevel algorithm in quenched QCD to study the correlation of singlet mesonic observables, which constitute the most demanding computations.
In particular, we discuss how we combine these advanced lattice techniques to estimate more efficiently disconnected diagrams.
\hfill \break

\section{Distillation}\label{sec:sec2}
We are interested in estimating two-point functions like
\begin{equation}
\langle \mathrm{O}(\vec{p}, t_1) \bar{\mathrm{O}}(\vec{p}, t_0)\rangle
=
\frac{1}{\mathcal{Z}}\int [dU] [dq] [d\bar{q}] e^{-S[q, \bar{q}, U]}\mathrm{O}(\vec{p}, t_1) \bar{\mathrm{O}}(\vec{p}, t_0),
\end{equation}
where $\mathrm{O}$ are interpolating operators with glueball quantum numbers,
$\mathcal{Z}$ is the QCD partition function and $S = S_f + S_g$ is the QCD action.
After integrating out the fermionic degrees of freedom, the correlation function in the quenched approximation reads
\begin{equation}\label{corr}
\langle \mathrm{O}(\vec{p}, t_1) \bar{\mathrm{O}}(\vec{p}, t_0)\rangle
=
\frac{1}{\mathcal{Z}} \int [dU]  e^{-S_g[U]} \langle \mathrm{O}(\vec{p}, t_1) \bar{\mathrm{O}}(\vec{p}, t_0)\rangle_F,
\end{equation}
where $\langle \mathrm{O}(\vec{p},t_1) \bar{\mathrm{O}}(\vec{p}, t_0)\rangle_F$ are Wick contractions expressed in terms of traces of products of Dirac propagators.
In particular, the interpolating operators that we consider in this work are singlet meson $\mathrm{O}_\Gamma$ interpolators projected to zero total momenta, 
whose expressions are given by
\begin{align}
\label{meson_op}
\mathrm{O}_\Gamma(\vec{0}, t_1)&=\sum_{\vec{x}} \bar{q}(x) \Gamma q(x) 
~\hspace{3cm} 
&&\text{with} \quad \Gamma=1, \gamma_5, \gamma_\mu, \gamma_4\gamma_5, \gamma_i\gamma_j ;
\end{align}
with $x=(\vec{x}, t_1)$.
The sum over spatial coordinates projects the operators to zero total momentum.
Given that the operators are all projected to zero momentum, we drop the momenta to simplify the notation.
The Wick contractions of these operators contain disconnected contributions which read
\begin{align}
\label{corrql}
\langle \mathrm{O}_\Gamma(t_1) \bar{\mathrm{O}}_\Gamma(t_0)\rangle_{F, \rm disc}
&=
\sum_{\vec{x}, \vec{y}}
\langle \mathrm{O}_\Gamma(x)\rangle_F \langle\bar{\mathrm{O}}_\Gamma(y)\rangle_{F},
\end{align}
where the explicit expressions of the Wick contractions contain quark loops, which read
\begin{align}
\mathcal{O}_\Gamma(x)
&:=
\langle \mathrm{O}_\Gamma(x)\rangle_F
=
\rm \mathrm{Tr} \left[D^{-1}(x,x) \Gamma \right].
\end{align}
In the distillation framework \cite{Peardon:2009d}, the Wick contractions in eq.~\eqref{corrql} are rewritten as
\begin{align}
\label{corrqldist}
&\langle \mathrm{O}_\Gamma(t_1) \bar{\mathrm{O}}_\Gamma(t_0)\rangle_{F, \rm disc} 
=
\mathrm{Tr} \left[\phi(t_1) \tau(t_1,t_1)\right] \mathrm{Tr}\left[\phi(t_0) \tau(t_0,t_0)\right]
\end{align}
where $\phi$ and $\tau$ are elementals and perambulators, respectively, whose expressions are
\begin{align}
\label{elem}
\phi_{ij}(t)_{\alpha \beta} &= \Gamma_{\alpha \beta} ~v_i(t)^\dagger v_j(t),
\\
\label{peramb}
\tau_{ij}(t_1,t_2)_{\alpha \beta}&=v_i(t_1)^\dagger D^{-1}(t_1,t_2)_{\alpha \beta}~ v_j(t_2).
\end{align}
In these expressions, $v(t)$ are the eigenvectors of the 3D Laplacian operator
\begin{equation}
\nabla^2(t)_{\vec{x}, \vec{y}} = 
-6 \delta_{\vec{x}, \vec{y}} 
+ \sum_{k=1}^{3} 
\left[ 
U_k(\vec{x}, t) \delta_{\vec{x} + \hat{k}, \vec{y}}	
+
U^\dagger_k(\vec{x}-\hat{k}, t) \delta_{\vec{x} - \hat{k}, \vec{y}}	
\right],
\end{equation}
which is constructed in terms of the gauge fields $U$. In this preliminary work, 
we consider only 10 eigenvectors and the subscripts $i,j$ run over the number of eigenvectors. 
To reduce UV fluctuations in the observables,
the gauge fields are appropriately smeared through APE smearing \cite{Falcioni:1984ei}.

\section{Two-level sampling for fermionic observables}
In the quenched approximation, the two-point functions in eq.~\eqref{corr} depend on the pure gauge action $S_g[U]$ and the Wick contractions
$\langle \mathrm{O}(t_1) \bar{\mathrm{O}}(t_0)\rangle_F$.
The pure gauge action $S_g[U]$ is local in the gauge fields as the action is constructed in terms of Wilson plaquettes.
However, the Wick contractions are expressed as traces of products of quark propagators, see for instance eq.~\eqref{corrql}, 
and the quark propagator depends on the values of the gauge fields over the full space-time.
In theories with a mass gap like QCD or 4D Yang-Mills, the physical signal of a correlation function decays exponentially with the distance
between the operators $\mathrm{O}(t_1)$ and $\bar{\mathrm{O}}(t_0)$.
Supported by empirical arguments \cite{Parisi:1983ae}, the quark propagator is suppressed on each gauge configuration according to 
\begin{equation}
|| D^{-1}(y,x)|| \sim e^{-\frac{1}{2}m_\pi |y-x|},
\end{equation}
where $m_\pi$ is the mass of the lightest pseudoscalar state and $||\sbullet[0.75]||$ is a gauge-invariant norm.
Therefore, contributions of the background gauge field configurations from points that are located far away 
from the operators decay exponentially with the distance.
Based on this exponential locality, an approximated quark propagator can be constructed to remove the dependence of the quark propagators from gauge fields located at distant points.
This approximation allows a two-level integration of the estimator in eq.~\eqref{corrql}, which reads
\begin{equation}\label{corr2lvl}
C(t_1-t_0) 
=
\frac{1}{\mathcal{Z}[U_B]} \int [dU_B]  e^{-S_g[U_B]} \bigl[ \mathcal{O}(t_1)\bigr] \bigl[\bar{\mathcal{O}}(t_0)\bigr].
\end{equation}
The two-level integration is possible when the operators are in different dynamical regions, say $t_0 \in \Lambda_0$, $t_1 \in \Lambda_2$.
Using the operators in eq.~\eqref{meson_op}, the local integrations $\left[ \sbullet[0.75]\right]$ 
of the quark loops read explicitly
\begin{align}
\left[\mathcal{O}_\Gamma(t_1)\right] &= \int [dU_1] ~\sum_{\vec{x}}\mathrm{Tr}\left[ D^{-1}_{\Omega_1}(x,x) \Gamma \right].
\end{align}
In the distillation framework, the local integration is performed for the elementals and perambulators, and the local integration of the quark loop at $t_1$ reads for instance
\begin{align}
\left[\mathcal{O}_\Gamma(t_1)\right] &= \int [dU_1] ~\sum_{\vec{x}}\mathrm{Tr} \left[\phi(x) \tau_{\Omega_1}(x,x)\right],
\end{align}
where the perambulators are computed using eq.~\eqref{peramb} with the approximated propagator $D^{-1}_{\Omega_r}$.
\begin{figure}[t!]
	\includegraphics[width=\textwidth]{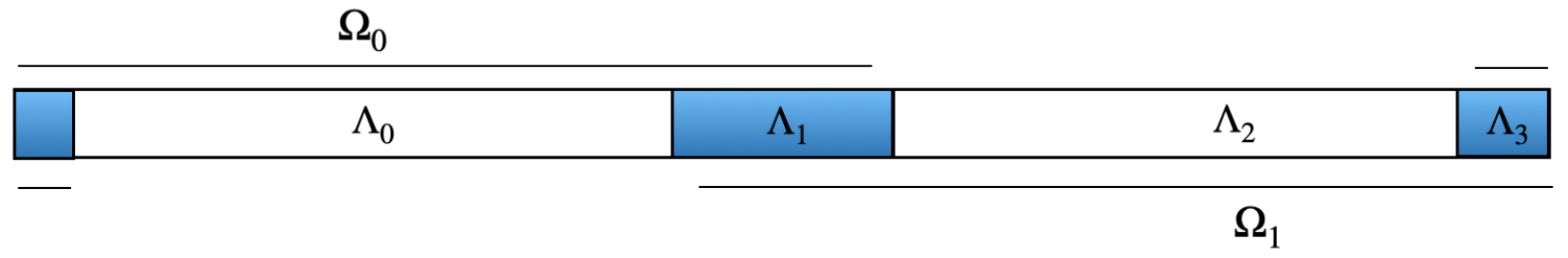}
	\caption{Domain decomposition used in this analysis. 
		The blue and white regions are the frozen and dynamical regions, respectively.} \label{Figure:dd}
	\end{figure}

In Fig.~\ref{Figure:dd}, we show the domain decomposition used in this work, 
where the temporal lattice extent is split in four regions:
$\Lambda = \Lambda_0 \oplus \Lambda_1 \oplus \Lambda_2 \oplus \Lambda_3$
and the Dirac propagators are computed in two overlapping regions:
$\Omega_0 = \Lambda_3 \oplus \Lambda_0 \oplus \Lambda_1$
and
$\Omega_1 = \Lambda_1 \oplus \Lambda_2 \oplus \Lambda_3$.
Notice that we use periodic boundary conditions. The details of the simulations are discussed in the next section.

\section{Numerical results}
\subsection{Details of the simulations}
We discretise a four dimensional $\rm SU(3)$ theory using the Wilson action with periodic boundary conditions.
The gauge configurations are generated at $\beta=6.0$ and with a volume $V/a^4=16^3\times 64$.
In order to perform a two-level integration, we use molecular dynamic integration to sample independent gauge configurations as discussed in \cite{Barca:2024fpc}. 
First, $N_0$ gauge configurations are generated by updating the gauge fields over the full space-time. 
These represent the fields $U_B$ over which we integrate in eq.~\eqref{corr2lvl}.
Second, for each of these stored gauge configurations, we generate a new trajectory of $N_1$ gauge fields according to the same probability distribution.
Along these new trajectories, the gauge fields are updated only in certain regions of the temporal extent, which are the regions $\Lambda_0$ and $\Lambda_2$.\footnote{The gauge configurations are well spaced along both levels to suppress any autocorrelation effects.}
Therefore, $\Lambda_0$ and $\Lambda_2$ are the dynamical regions, while $\Lambda_1$ and $\Lambda_3$ are the frozen regions.
In the simulation, the frozen regions have $t/a\in \{0,1,61,62,63\}$ for $\Lambda_3$ and $t/a\in\{29,30,31,32,33\}$ for $\Lambda_1$, 
while the remaining lattice sites compose $\Lambda_0$ and $\Lambda_2$.

We generate $N_0=101$ level-0 gauge configurations and $N_1 = 200$ level-1 (local) updates and we label these gauge configurations $U^{(ij)}$,
where the superscripts $i,j$ refer to the $i$-th level-0 and $j$-th level-1 gauge configuration $U^{(ij)}$, respectively.
For each of the $N_0 \times N_1$ gauge configurations, we compute the full and approximated Wilson Dirac propagator
using the Wilson clover action with $\kappa=0.13393$, where the lightest pseudoscalar state has a mass $m_\pi\approx 760~\rm MeV$.
The quark mass was tuned such that the lowest non-interacting $\pi\pi$ energy is very close to the pure gauge scalar glueball,
which at a similar scale ($\beta=5.99$) is $m_G^{0^{++}}\approx 1560~\rm MeV$ according to \cite{Athenodorou:2020ani}.

To compute the approximated quark propagator we adopt a domain decomposition method \cite{Luscher:2003qa} and impose Dirichlet boundary conditions.
For instance, the quark propagator $D^{-1}_{\Omega_0, ij}(x,x)$ is computed on the configuration $U^{(ij)}$ in the region $\Omega_0$, 
neglecting contributions from $\Lambda_2$, see Fig.~\ref{Figure:dd}.

\subsection{Analysis of disconnected 2-points functions with 1- and 2-level estimators}
\begin{figure}[t!]
\centering
\includegraphics[width=0.95\textwidth]{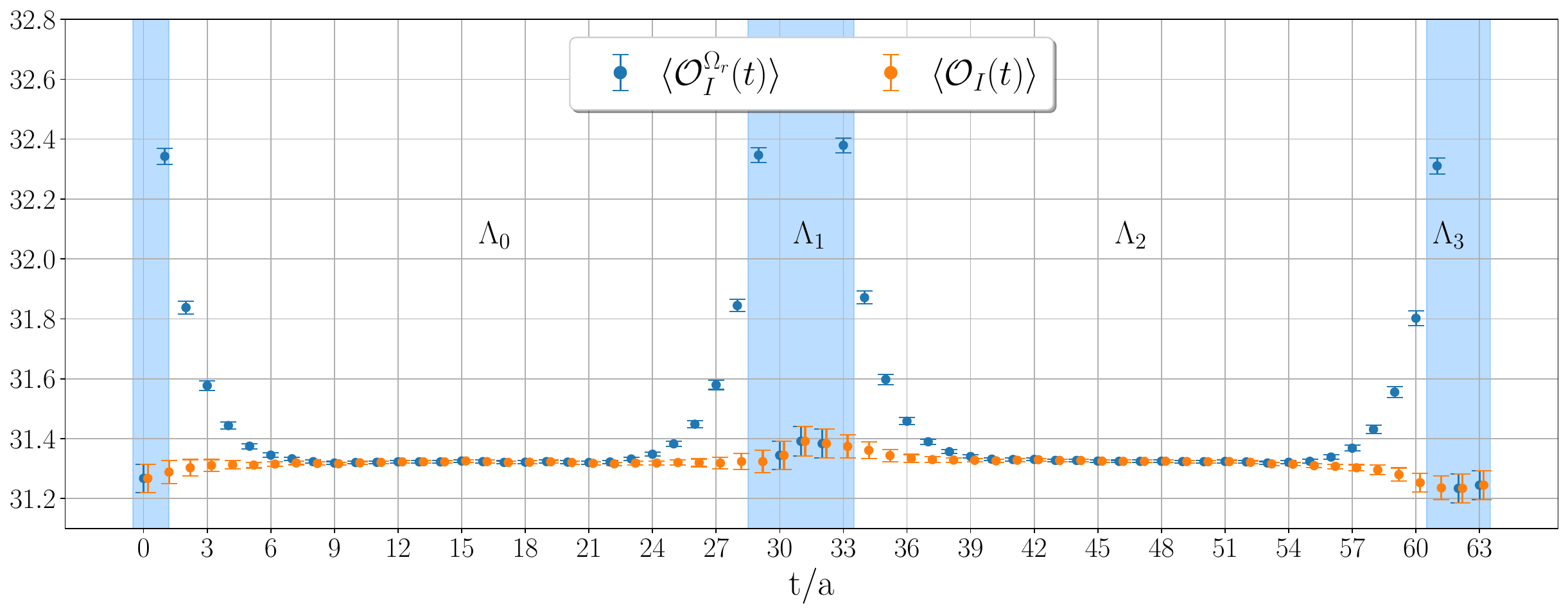}
\caption{Comparison between the 1-point functions of scalar quark loops using the approximated (blue) and full (orange) quark propagator.
The scalar channel has non-vanishing expectation value because it has vacuum quantum numbers.
The blue vertical bands highlight the location of the frozen regions $\Lambda_1$ and $\Lambda_3$.
These estimates are computed using eqs.~\eqref{1pt_std}-\eqref{1pt_approx} with $i=1, ..., N_0$ and $j=1, ..., N_1$, and the errors
are estimated on the level-0 configurations using the $\Gamma$-method.
}\label{Figure:1pt}
\end{figure}

Using the $N_0\times N_1$ gauge configurations, which we store on disk, 
we compute the traces of quark loops using the full and approximated Wilson Dirac propagators:
\begin{align}
\label{1pt_std}
\mathcal{O}_{\Gamma, ij}(t_1) &= \sum_{\vec{x}}\mathrm{Tr}\left[D^{-1}_{ij}(x,x) \Gamma \right]
\\
\label{1pt_approx}
\mathcal{O}^{\Omega_1}_{\Gamma, ij}(t_1) &= \sum_{\vec{x}}\mathrm{Tr}\left[D^{-1}_{\Omega_1, ij}(x,x) \Gamma \right],
\hspace{1.5cm}
\bar{\mathcal{O}}^{\Omega_0}_{\Gamma, ij}(t_0) = \sum_{\vec{y}}\mathrm{Tr}\left[D^{-1}_{\Omega_0, ij}(y,y) \Gamma \right],
\hspace{0.5cm}
\end{align}
with $x=(\vec{x}, t_1)$ and $y=(\vec{y}, t_0)$. In particular, we use distillation to estimate the traces.
In Fig.~\ref{Figure:1pt}, we show a comparison between scalar quark loops using the full and approximated quark propagators
in eqs.~\eqref{1pt_std}, \eqref{1pt_approx}, respectively.
The scalar quark loops computed with the full propagator (orange) fluctuate within the errors around a constant value.
The data points inside $\Lambda_1$ and $\Lambda_3$ are noisier because the gauge fields are not updated in these regions.
Notice that the orange data points are shifted along the x-axis to increase visibility.
The scalar quark loops computed with the approximated quark propagator (blue) fluctuate also within the errors around 
the same constant value when the quark loops are sufficiently distant from the other dynamical region.
The approximation gets better with the distance from the other dynamical region, as expected from the exponential locality of the quark propagator.
Near the boundaries, and therefore close to the other dynamical region, there are visible effects due to the Dirichlet boundary conditions.
These effects are then taken into account for the final two-level estimator of the two-point functions.
Inside the internal region of the boundaries ($t/a=0,30,31,32, 62, 63$), we use the full propagator in both cases,
which explains why the quark loops coincide in these sites.

Using the quark loops computed with the full propagator, we estimate the disconnected two-point functions with the traditional sampling as
\begin{equation}
\label{c2pt_std}
C^{\textnormal{1-lvl}}_{\Gamma}(t_1, t_0)
=
\frac{1}{N_0 N_1}\sum_{ij} \mathcal{O}_{\Gamma, ij}(t_1) \bar{\mathcal{O}}_{\Gamma, ij}(t_0).
\end{equation}
The superscript "1-lvl" refers to the fact that the operators are correlated both along level-0 and level-1, 
i.e., treating the gauge configurations along the two levels as a single trajectory.
Using the observables computed with the approximated propagator instead, 
we estimate the disconnected two-point functions using both 1-level and 2-level integrations, respectively
\begin{align}
\label{c2pt_1lvl_approx}
\widetilde{C}^{\textnormal{1-lvl}}_{\Gamma}(t_1, t_0)
&=
\frac{1}{N_0 N_1}\sum_{ij} \mathcal{O}^{\Omega_1}_{\Gamma, ij}(t_1) \bar{\mathcal{O}}^{\Omega_0}_{\Gamma, ij}(t_0),
\\
\label{c2pt_2lvl_approx}
\widetilde{C}^{\textnormal{2-lvl}}_{\Gamma}(t_1, t_0)
&=
\frac{1}{N_0}\sum_{i} 
\left[\frac{1}{N_1}\sum_j \mathcal{O}^{\Omega_1}_{\Gamma, ij}(t_1)\right] 
\left[\frac{1}{N_1}\sum_k \bar{\mathcal{O}}^{\Omega_0}_{\Gamma, ik}(t_0)\right].
\end{align}
The difference between eq.~\eqref{c2pt_std} and eq.~\eqref{c2pt_1lvl_approx} is the approximation adopted
for the quark propagator in the latter. Therefore, the correction due to this approximation can be estimated as
\begin{equation}
\label{correction}
\delta(t_1, t_0) = C^{\textnormal{1-lvl}}_{\Gamma}(t_1, t_0) - \widetilde{C}_{\Gamma}^{\textnormal{1-lvl}}(t_1, t_0).
\end{equation}
This correction is then added to the (improved) two-level estimator in eq.~\eqref{c2pt_2lvl_approx}
\begin{equation}
\label{c2pt_2lvl_corrected}
C^{\textnormal{2-lvl}}_{\Gamma}(t_1, t_0) = \widetilde{C}^{\textnormal{2-lvl}}_{\Gamma}(t_1, t_0) + \delta(t_1, t_0)
\end{equation}
to obtain a corrected two-level estimator for the two-point functions.
For the error analysis, we use the $\Gamma$-method \cite{Wolff:2003sm} to compute the errors on the level-0 configurations.

\begin{figure}[t!]
\includegraphics[width=1.\textwidth]{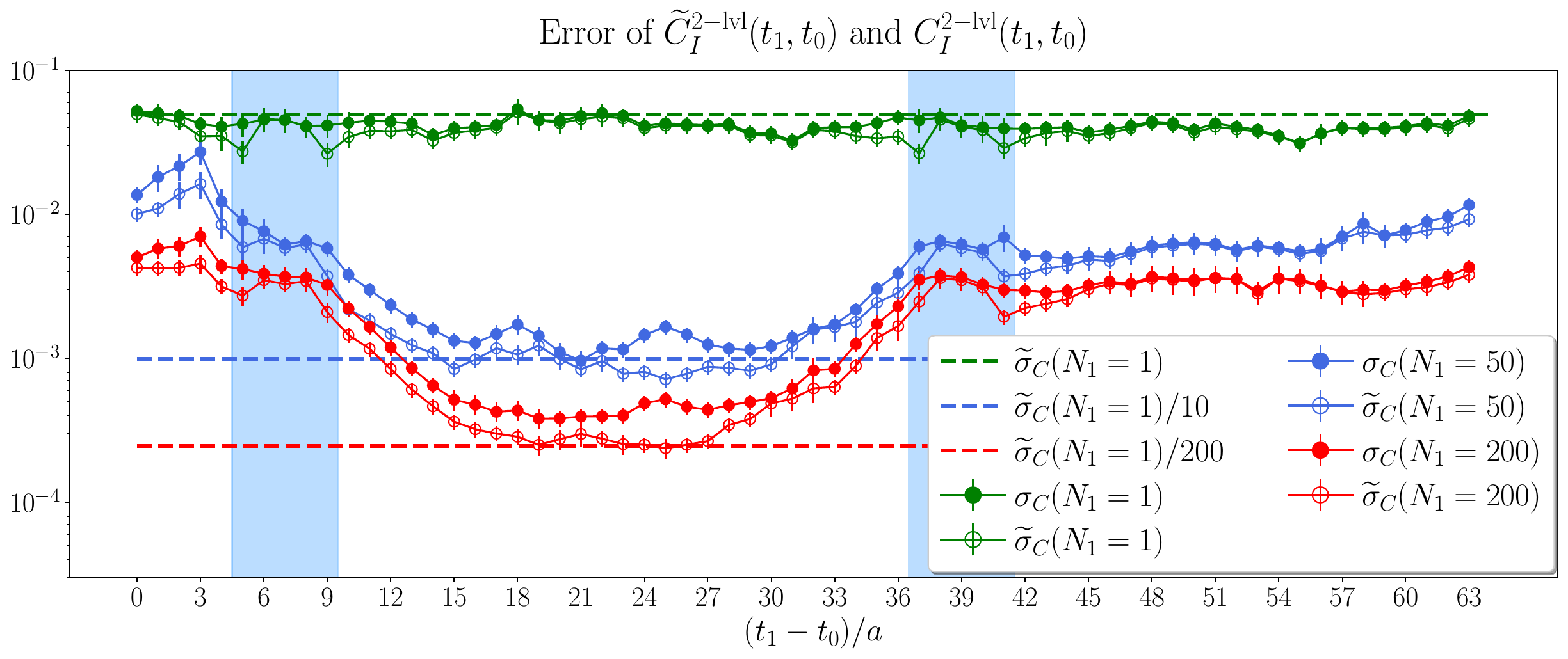}
\caption{Statistical errors of the two-level estimators for the disconnected scalar two-point functions before (empty circles)
	and after (filled circles) adding the correction in eq.~\eqref{correction}.
	The empty (filled) circles correspond to the error of the estimator in eq.~\eqref{c2pt_2lvl_approx} (\eqref{c2pt_2lvl_corrected}) in the scalar channel.
}
\label{Figure:error_reduction_scalarql}
\end{figure}
In Fig. \ref{Figure:error_reduction_scalarql}, we show the statistical error for the scalar disconnected 2-point functions $C_I(t)$
using the approximated and corrected two-level estimators, respectively in eqs.~\eqref{c2pt_2lvl_approx} and \eqref{c2pt_2lvl_corrected}.
While the number of level-0 configurations is fixed to $N_0=101$ and one quark loop is located at $t_0=24a$, 
we vary the number of submeasurements and the location of the other quark loop.
We do not observe a significant bias introduced by the addition of the correction term, at least for this particular observable.
When the quark loops are in the same region, the error is fluctuating around a constant value, which decreases like the standard sampling ($1/\sqrt{N_1}$); 
while when the quark loops are in two different dynamical regions the error is decreasing as observed in the pure gauge analysis in \cite{Barca:2024fpc}.
In particular, the variance scales with $1/N_1^2$, with additional corrections that decrease exponentially with the distance from the frozen regions.

\begin{figure}[t
	!]
	\includegraphics[width=\textwidth]{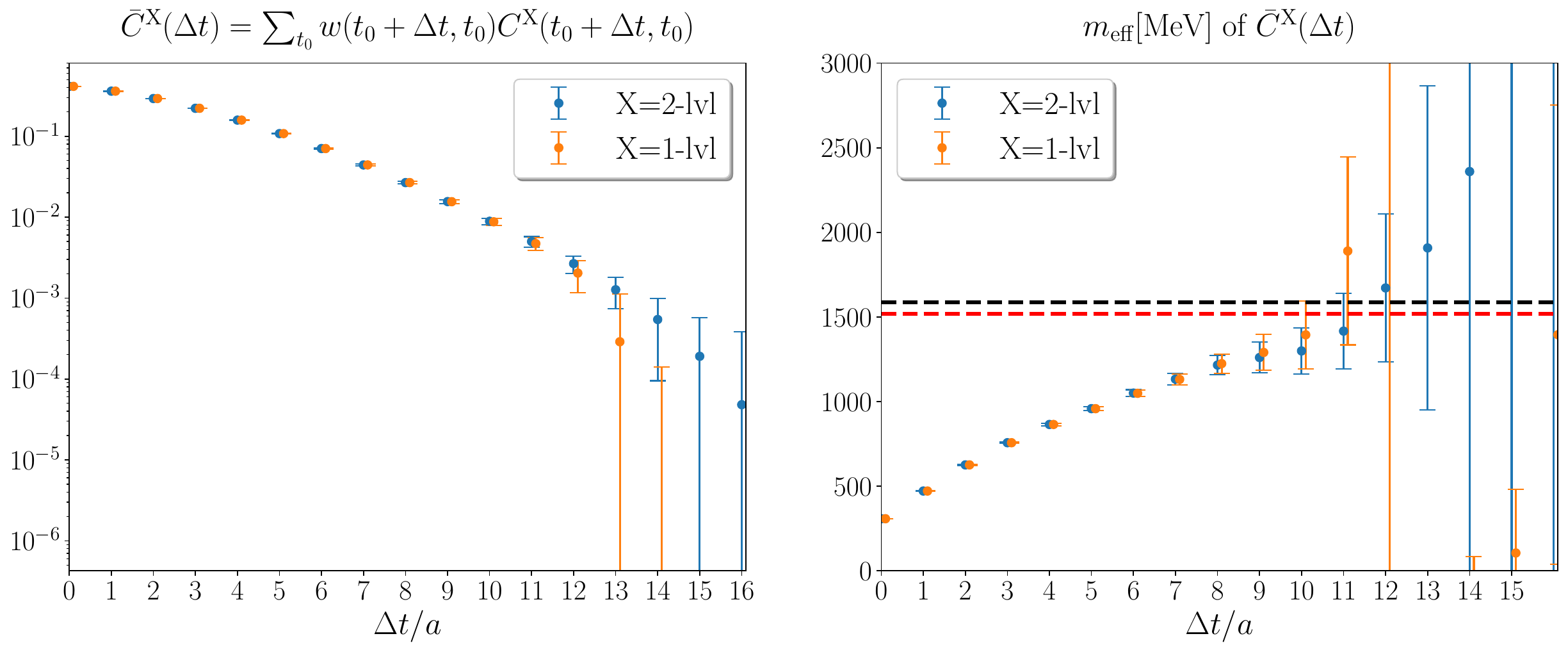}
	\caption{(left) Comparison between a 1-level \eqref{c2pt_std} and 2-level \eqref{c2pt_2lvl_corrected} estimator
		for the weighted average disconnected two-point functions of scalar quark loops;
		(right) Comparison between the two sampling techniques for the effective masses of scalar disconnected two-point functions.
		The red dashed lines correspond to the lowest non-interacting $\pi\pi$ energy, 
		while the black dashed lines represent the mass of the lowest scalar glueball in pure gauge theory at $\beta=5.99$ \cite{Athenodorou:2020ani}.
	}
	\label{Figure:c2pt_meff_scalarql}
\end{figure}

\paragraph{Weighted average}
Given that the correlators at different temporal source positions $t_0$ fluctuate differently at fixed $\Delta t = t_1-t_0$,
especially with the two-level integration, we construct a weighted average correlator as in \cite{Barca:2024fpc}, which reads
\begin{equation}
\bar{C}^{\rm X}(\Delta t) = \sum_{t_0} w(t_0+\Delta t, t_0) C^{\rm X}(t_0+\Delta t, t_0)
\hspace{2cm}
\rm \text{with} \quad X=\textnormal{1-lvl, ~2-lvl}.
\end{equation}
The weight functions are chosen to be proportional to the inverse of the variance, 
i.e. $w(t_1, t_0) = \mathcal{N} \sigma^2(t_1, t_0)^{-1}$, 
and the normalisation $\mathcal{N}$ is chosen such that $\sum_{t_0} w(t_1, t_0)=1 ~ \forall~ t_1$.
In Fig.~\ref{Figure:c2pt_meff_scalarql}, we show a comparison between the standard and two-level estimators
of the disconnected two-point correlation functions and their effective masses, using the scalar meson operator. 
The statistical estimators for the 1-level and 2-level sampling are constructed using the same statistics, i.e. $N_0\times N_1 = 101\times 200$.
With the two-level sampling, the signal survives a few more time slices, but the statistics is still not enough to extract a reliable
estimate of its effective mass, see right plot. 
Please note that this is the signal of the disconnected contribution only.
In order to extract the energy of the scalar meson/glueball state, we will include the connected contributions, 
which we compute with a 1-level integration as they do not suffer from a severe S/N ratio.
If we are only interested in the disconnected signal, increasing the submeasurements by a further factor of $5$,
and thus reducing the error by a similar factor for $t>11a$, would help to identify a plateau.
As a comparison, we add anyways in the right plot of Fig.~\ref{Figure:c2pt_meff_scalarql} the values of the non-interacting $\pi\pi$ energy and the scalar glueball mass at a similar scale in pure gauge theory.

\section{Conclusions}
In this work, we have combined, for the first time, distillation techniques with a two-level 
sampling algorithm to study two-point functions of quark loops.
In order to use the two-level sampling, a factorisation of the quark propagator is adopted, 
and the quark loops are submeasured in different regions.
We report results for the scalar two-point functions with a two-level algorithm
and we find that the error decreases as expected from our previous study in pure gauge theory.
In particular, we find that the two-level integration reduces substantially the statistical error of the disconnected contributions,
as showcased in Fig.~\ref{Figure:error_reduction_scalarql}. 
This also results in an error reduction in the effective mass of the disconnected piece for the would-be $f_0$ state, 
where the signal is extended for longer distances, see Fig.~\ref{Figure:c2pt_meff_scalarql}.
We conclude that a two-level algorithm is anyways a superior method to study disconnected observables,
and will be applied for future studies of glueballs in full QCD.

\paragraph{Acknowledgements}
The work is supported by the German Research Foundation (DFG) research unit 
FOR5269 "Future methods for studying confined gluons in QCD".
Simulations were performed on JUWELS in Jülich and PAX in Zeuthen.
J.F. and J.U.-N. are also supported by the Inno4scale project, 
which received funding from the European High-Performance Computing 
Joint Undertaking (JU) under Grant Agreement No. 101118139.

\bibliography{bibliography}
\bibliographystyle{JHEP}

\end{document}